\documentstyle[sprocl,epsfig]{article}

\def\ra{\rightarrow}

\def\beq{\begin{equation}}
\def\eeq{\end{equation}}
\def\bea{\begin{eqnarray}}
\def\eea{\end{eqnarray}}
\newcommand{\lsim}{\raisebox{-0.13cm}{~\shortstack{$<$ \\[-0.07cm] $\sim$}}~}
\newcommand{\gsim}{\raisebox{-0.13cm}{~\shortstack{$>$ \\[-0.07cm] $\sim$}}~}
\newcommand{\ee}{e^+e^-}

\newcommand{\tb}{\tan\beta}

\newcommand{\dt}{{\rm d}t}

\def\t1{\tilde{t_1}}

\begin{document}
\vspace{-2truecm}
\begin{flushright}
PM/99--37
\end{flushright}

\title{STOP--STOP--HIGGS PRODUCTION\\ 
AT FUTURE LINEAR COLLIDER} 

\author{A. DJOUADI, J.-L. KNEUR~\footnote{Speaker}  and G. MOULTAKA }

\address{Physique Math\'ematique et Th\'eorique, UMR No 5825--CNRS, \\
Universit\'e Montpellier II, F--34095 Montpellier Cedex 5, France}


\maketitle\abstracts{ In the Minimal Supersymmetric Standard Model, the cross
section for the associated production of the lightest neutral Higgs boson with
the lightest top squark pairs can be rather substantial at high energies. We
summarize the properties of this production process at a future $e^+e^-$ linear
collider, including the $\gamma \gamma$ mode. } 
\section{Introduction} 
In the Minimal Supersymmetric Standard Model (MSSM)~\cite{R1}, if the mixing
between third generation squarks is large, stops/sbottoms can be rather
light~\cite{qmix} and at the same time, their coupling to Higgs bosons can 
become substantial.  This might have a rather large impact on the phenomenology
of the MSSM Higgs bosons~$^{\rm 3-7}$. More precisely, concentrating on 
scalar top quarks, their mass eigenvalues are given by
\beq
m_{\tilde{t}_{1,2}}^2 = m_t^2 + \frac{1}{2} \left[ m_{\tilde Q_L}^2 + m_{\tilde
t_R}^2 +\cdots \mp \sqrt{ (m_{\tilde Q_L}^2 - m_{\tilde t_R}^2 +\cdots)^2 + 
4m_t^2 \tilde{A}_t^2 } \right]\;, 
\eeq
where $m_{\tilde Q_L}$, $m_{\tilde t_R}$ are the soft-SUSY breaking scalar 
masses and the dots stand for the $D$--terms $\propto M^2_Z \cos 2\beta$.  In the 
decoupling limit (the lightest $h$ boson is Standard Model like and the other 
bosons $A,H$ and $H^\pm$, are very heavy), the expressions of the coupling $h
\tilde{t_1} \tilde{t_1}$ simply reads ($\theta_t$ is the mixing angle and $s_W
\equiv \sin\theta_W$)
\begin{equation}
g_{h \tilde{t}_1 \tilde{t}_1 } = \cos 2\beta \left[ \frac{1}{2} \cos^2 \theta_t
- \frac{2}{3} s^2_W \cos 2 \theta_t \right] + \frac{m_t^2}{M_Z^2} + \frac{1}{2}
\sin 2\theta_t \frac{m_t \tilde{A}_t } {M_Z^2}\;.  \label{ghtt}
\end{equation}
Large values of $\tilde{A}_t\equiv A_t -\mu/ \tan\beta$ lead to 
$g_{h \tilde{t}_1 \tilde{t}_1} \sim \tilde{A}_t$  
and to an almost maximal $\tilde{t}$ 
mixing angle, $|\sin 2 \theta_t| \simeq 1$, in particular
if $m_{\tilde Q_L} \simeq m_{\tilde t_R}$.  
The measurement of this important coupling would open a
window to probe directly some of the soft--SUSY breaking terms of the
potential. To measure Higgs--squarks couplings directly, one needs to consider
the three--body associated production of Higgs bosons with scalar quark
pairs~$^{\rm 4-7}$. This is the supersymmetric analog to the processes
of Higgs boson radiation from top quark lines~\cite{ppttH,eetth} which allows 
to probe the $t\bar{t}$--Higgs Yukawa coupling directly. [At the LHC, the 
process $pp \to t\bar{t}+$Higgs~\cite{ppttH} although not competitive with the 
gluon fusion mechanism~\cite{R7}, can provide a complementary signal since 
backgrounds are smaller.] 

Here, we report on the production of a light Higgs boson $h$ in association
with top squarks at future $e^+e^-$ linear machines~\cite{Xtth}, including the
$\gamma \gamma$ option, both in the unconstrained MSSM and minimal SUGRA cases.
For simplicity, we work in the approximation of being close to the decoupling 
limit, which  implies that we do not consider other Higgs production processes 
with e.g. the heavy $H$ or $A$ bosons produced in association 
with top squarks. For large masses, these processes will be suppressed by 
phase--space well before the decoupling regime is reached.
\section{Associated production at e$^+$e$^-$ colliders} 
At future linear $\ee$ colliders, the final state $\tilde{t}_1 \tilde{t}_1 h$
may be generated in three ways: $(i)$ two--body production of a mixed pair
of top squarks and the decay of the heaviest stop to the lightest one and a 
Higgs boson, $(ii)$ the continuum production in $\ee$ annihilation $\ee \ra
\tilde{t}_1 \tilde{t}_1h$ and $(iii)$  the continuum production in 
$\gamma \gamma$ collisions $\gamma \gamma \ra \tilde{t}_1 \tilde{t}_1h$.

\vspace*{-4mm}

\subsection{Two--body production and decay} 
The total cross section~\cite{R8} for the process $(i)$ should, in principle, 
be large enough for the final state to be copiously produced. However, $\sigma( 
\ee \ra \tilde{t}_1 \tilde{t}_2$) involves the $Z\tilde{t}_1 \tilde{t}_2$ 
coupling, proportional to $\sin 2 \theta_t$, while the $\vert g_{h \tilde{t}_1 
\tilde{t}_2}\vert $ coupling  in most of the parameter space is proportional 
to $\cos 2\theta_t$, such that the cross section times branching ratio will be 
very small in the no mixing [$\theta_t \sim 0$] and maximal mixing $[|\theta_t| 
\sim \pi/4]$ cases. 
[In addition, the decay width $\tilde{t}_2 \ra h\tilde{t}_1$ is 
in general much smaller~\cite{R8} than the $\tilde{t}_2$ decay widths into 
chargino and neutralinos]. Nevertheless, there are regions of the MSSM 
parameter space where
the combination $\sin 2 \theta_t \times \cos 2 \theta_t$ can be large; this
occurs typically for a not too small $m_{\tilde t_L}$--$m_{\tilde t_R}$ 
splitting and moderate $\tilde A_t$ values.\cite{Xtth} In this case, which is 
often realized in the mSUGRA scenario, this mechanism is visible for the 
expected high luminosities\cite{TESLA} $\int {\cal L}\dt \sim 500$ fb$^{-1}$.  

This is illustrated in Fig.~1, where the cross section $\ee \ra \tilde{t}_1 
\tilde{t}_2$ times the branching ratio BR($\tilde{t}_2 \ra \tilde{t}_1 h)$ is 
shown as a function of the $\tilde{t}_1$ mass at a c.m. energy of $\sqrt{s}
=800$ GeV. We have chosen a mSUGRA scenario with $\tan\beta=30$, $m_{1/2} =$ 
100 GeV, $A_0 = -600$ GeV and sign$(\mu)= +$. (The dotted lines show the 
contribution of the  non--resonant process, discussed below, for the same 
input choice). The cross section can reach the level of 1 fb for relatively
small $m_{\tilde{t}_1}$ values, leading to thousand events in a few years, for 
$\int {\cal L}\dt \sim 500$ fb$^{-1}$. 
\begin{figure}[htb]
\vspace{-.5cm}
\begin{center}
\mbox{
\psfig{figure=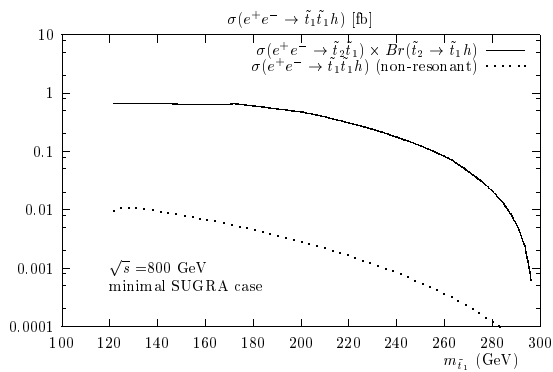,width=10cm}}
\end{center}
\vspace{-9.cm}
\caption[]{The production cross section $\sigma (e^+e^- \ra \tilde{t}_1 
\tilde{t}_1 h$) [in fb] as a function of $m_{\tilde{t}_1}$ in the mSUGRA 
case; $\tan\beta=$ 30, $m_{1/2} =$ 100 GeV and $A_0 =-600$ GeV.}
\end{figure}
\subsection{Production in the continuum in e$^+$e$^-$ collisions}
The cross section for the process $\ee \ra \tilde{q}_i \tilde{q}_i \Phi$ with 
$\Phi$ the CP--even Higgs boson $h$ or $H$, and $\tilde{q}_i$ any of the two 
squarks has been calculated in \cite{Xtth}.
We show in Fig.~2 the rates for the $\tilde{t}_1 \tilde{t}_1 h$ final state  as
a function of the $\tilde{t}_1$ mass in the unconstrained MSSM, at
$\sqrt{s}=800$ GeV.  For not too large $\tilde{t}_1$ masses and large values of
the parameter $\tilde A_t$, the production cross sections can exceed  
$1$ fb, to be compared to the
SM--like process $\ee \ra t \bar{t}h$ \cite{eetth} of the order of 2 fb for
$M_h \sim 130$ GeV.  This provides more than one thousand events in a few
years, with a luminosity  $\int {\cal L}\dt \sim 500$ fb$^{-1}$, which should
be sufficient to isolate the final state and measure $g_{\tilde{t}_1
\tilde{t}_1 h}$  with some accuracy.  
\begin{figure}[htbp]
\vspace{-5mm}
\begin{center}
\mbox{
\psfig{figure=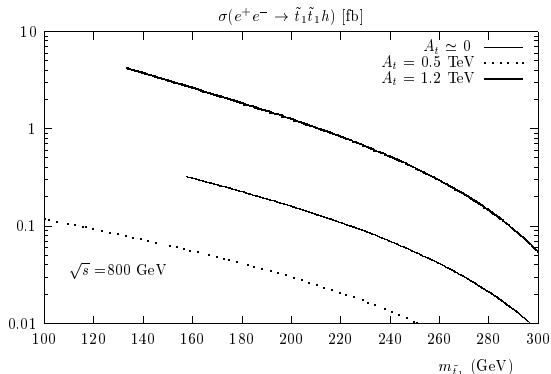,width=10cm}}
\end{center}
\vspace*{-9cm}
\caption[]{The cross section $\sigma (e^+e^- \ra \tilde{t}_1 \tilde{t}_1 h$) 
[in fb] as a function of the $\tilde{t}_1$ mass and the choices $\tb=30$, 
$A_t=$ 0 (0.5) TeV; $\tb=3$, $A_t=$ 1.2 TeV (and $\mu =-$600 GeV).} 
\end{figure}

Note however that $\tilde A_t$ cannot be arbitrarily large without conflicting
present constraints: more precisely, the absence of charge and color breaking
minima (CCB) \cite{CCB} can put rather stringent bounds on $\tilde{A}_t$, and
large $\tilde{A}_t$ values also generate potentially large contributions to
electroweak high--precision observables, in particular to the $\rho$ parameter
\cite{drho}, severely constrained  by LEP1 data.\cite{LEPrho} 
Those constraints, as well as the present 
experimental lower bounds on the top squark and Higgs 
boson masses, were systematically taken into account in our analysis.
We observed
that in the unconstrained MSSM case, the continuum production cross section in
$\ee$ annihilation $\sigma(\ee \ra \tilde{t}_1 \tilde{t}_1 h$) is often larger
than the resonant cross section for the production of $\tilde{t}_1 \tilde{t}_2$
and the subsequent 2--body decay $\tilde{t}_2 \ra \tilde{t}_1 h$, but this is
not generic. Indeed, in a situation where both a non-negligible $m_{\tilde
t_L}$--$m_{\tilde t_R}$ splitting and a moderate $\tilde A_t$ occurs, provided
there is sufficient phase space allowed, the production via a resonant $\tilde
t_2$ becomes competitive and even dominant, as illustrated in Fig. 1.

\vspace*{-4mm}

\subsection{Production in the continuum in $\gamma \gamma$ collisions} 
Future high--energy $\ee$ linear colliders can be turned into high--energy 
$\gamma \gamma$ colliders, with the high energy photons coming from Compton
back--scattering of laser beams.\cite{laser} The c.m. energy of the $\gamma 
\gamma$ collider is expected to be as much as $\sim 80\%$ of the one of the 
original $\ee$ machine. However, the total luminosity is expected to be 
somewhat smaller than the one of the $\ee$ mode.
\begin{figure}[htb]
\begin{center}
\vspace{-.5cm}
\mbox{
\psfig{figure=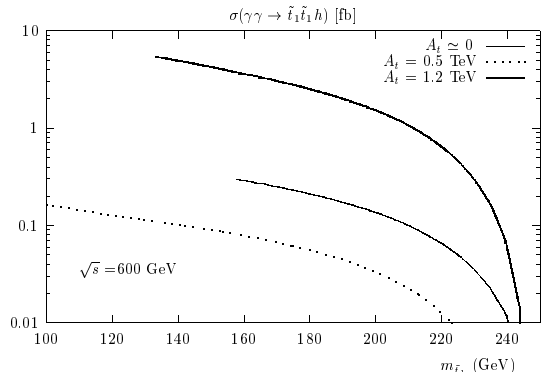,width=10cm}}
\end{center}
\vspace{-9cm}
\caption[]{$\sigma (\gamma\gamma \ra \tilde{t}_1 
\tilde{t}_1 h$) [in fb] at $\sqrt{s_{\gamma \gamma}} =600$ GeV as a function 
of $m_{\tilde{t}_1}$. The other parameters have the same values as in Fig.~2.}
\end{figure}
The total cross section for the subprocess $\gamma \gamma \ra \tilde{t}_1
\tilde{t}_1 h$, calculated in \cite{Xtth}, is shown in Fig.~3 at a 
two--photon c.m. energy $\sqrt{s}_{\gamma \gamma}  
\lsim 0.8 \sqrt{s}_{ee} =600$ GeV and as a function of the
$\tilde{t}_1$ mass, without convolution with the photon spectrum and with the
same inputs and assumptions as in Fig.~2 to compare with the $\ee$ mode. 
Because the c.m. energy of the $\gamma \gamma$ collider is only $\sim 80\%$ of
the one of the original $\ee$ machine, the process is of course less
phase--space favored than in the $\ee$ mode.  Nevertheless, the cross section
for the $\tilde{t}_1 \tilde{t}_1h$ final state is of the same order as in the
$\ee$ mode for c.m. energies not too close to the kinematical threshold, and
the process might be useful to obtain complementary information since it does
not involve the $Z$--boson and $\tilde{t}_2$ exchanges.  If the luminosities of
the $\gamma \gamma$ and $\ee$ colliders are comparable, a large number of
events might be collected for small stop masses and large 
$\tilde A_t$ values.  

\vspace*{-4mm}

\subsection{Decay modes and signal}
Top squarks in the mass range discussed above will mainly decay~\cite{R8} into 
a $c$ quark + neutralino, $\tilde{t}_1 \to c\chi_1^0$, or a $b$ quark + 
chargino,
$\tilde{t}_1 \ra b\chi^+$.  In this latter case the lightest chargino,
$\chi_1^+$  will decay into the LSP and a real or virtual $W$ boson, leading to
the same topology as in the case of the top quark decay, but with a large
amount of missing energy due to the undetected LSP
(three and four--body decays~\cite{fourbody} of the stop are also
possible with the same topology). At $\ee$ colliders one can
use the dominant decay mode of the lightest Higgs boson, $h \ra b\bar{b}$. The
final state topology will then consist of $4b$ quarks, two of them peaking at
an invariant mass $M_h$, two real (or virtual) $W$'s and missing energy. With
efficient micro--vertex detectors, this final state should be rather easy to 
detect.  

\vspace*{-4mm}

\section{Conclusions} 
At $\ee$ colliders with c.m. energies $\sqrt{s} \gsim 500$ GeV and with very
high luminosities $\int {\cal L}\dt \sim 500$ fb$^{-1}$, the process $\ee \ra
\tilde{t}_1 \tilde{t}_1h$ can lead to several hundreds of events, since the
cross sections can exceed the level of a 1 fb for not too heavy top squarks and
large trilinear coupling, $\tilde A_t \gsim 1$ TeV.  In the case where the top
squark decays into a $b$ quark and a real/virtual chargino, the final state
topology with $4b$ quarks, missing energy and additional jets or leptons will
be rather spectacular and should be easy to be seen experimentally, thanks to
the clean environment of these colliders.  In the $\gamma \gamma$ option of the
$\ee$ collider, the cross sections are similar as previously far from the
particle thresholds, but are suppressed for larger masses because of the
reduced c.m. energies; for $\gamma \gamma$ luminosities of the same order as
the original $\ee$ luminosities, the $\tilde{t}_1 \tilde{t}_1h$ final state
should also be observable in this mode, at least in some areas of the MSSM 
parameter space.  

The production cross section of the $\tilde{t}_1 \tilde{t}_1 h$ final state 
is directly proportional to the square of the $\tilde{t}_1 \tilde{t}_1 h$
couplings, therefore studying this process will allow to measure this important
coupling and to probe directly some of the soft--SUSY breaking parameters.  

\vspace*{-4mm}

\section*{References}

\end{document}